\documentclass[preprint]{aastex}

\newcommand{\etal}{{\it{et al.}}~}
\newcommand{\ie}{{\it{i.e.}}~}
\newcommand{\eg}{{\it{e.g.}}}

\def\ltaprx{\, \buildrel < \over \sim \,}

\begin{document}

\title{Stellar Populations in the Phoenix Dwarf (dIrr/dSph) Galaxy
as Observed by HST/WFPC2
\footnote{Based on observations with the NASA/ESA {\it Hubble Space
Telescope}, obtained at the Space Telescope Science Institute, operated
by AURA Inc under contract to NASA}}

\author{
Jon~A.~Holtzman\altaffilmark{1},
Graeme~H.~Smith\altaffilmark{2}, and
Carl~Grillmair\altaffilmark{3}
}
\altaffiltext{1}{Department of Astronomy, New Mexico State
University, Dept 4500
Box 30001, Las Cruces, NM 88003, holtz@nmsu.edu}
\altaffiltext{3}{UCO/Lick Observatory, University of California,
 Santa Cruz, CA 95064, graeme@ucolick.org}
\altaffiltext{3}{SIRTF Science Center, M/S 100-22, California
Institute of Technology, 770 S. Wilson, Pasadena, CA 91125,
carl@ipac.caltech.edu}

\begin{abstract}

We present HST/WFPC2 photometry of the central regions of the Phoenix dwarf.
Accurate photometry allows us to: 1) confirm the existence of the
horizontal branch previously detected by ground-based observations, and
use it to determine a distance to Phoenix, 2) clearly detect the
existence of multiple ages in the stellar population of Phoenix, 3) determine
a mean metallicity of the old red giant branch stars in Phoenix,
and suggest that Phoenix has evolved chemically over its
lifetime, 4) extract a rough star formation history for the central
regions which suggests that Phoenix has been forming stars roughly
continuously over its entire lifetime.

\end{abstract}


\section{Introduction}

The Phoenix dwarf galaxy, discovered by Schuster \& West (1976), has a
morphological type of dIrr/dSph (Mateo 1998), meaning that although
this galaxy has a low surface brightness and a mass and morphology like
a dwarf spheroidal (dSph) galaxy, there is evidence of recent star
formation (Canterna \& Flower 1977). Color-magnitude diagrams obtained
from ground-based images have been described by Ortolani \& Gratton
(1988), van de Rydt, Demers, \& Kunkel (1991), Held, Saviane, \& Momany
(1999), and Martinez-Delgado, Gallart, \& Aparicio (1999). These studies
find a broadly distributed metal-poor population of evolved giants, as
well as a concentration of young stars in the central region of the
galaxy. Several studies have found H~I in the direction of Phoenix
(Carignan, Demers, \& Cote 1991; Young \& Lo 1997); most recently,
St-Germain, Carignan, Cote, \& Oosterloo (1999) have found that
a mass of $\sim 10^5$ $M_{\odot}$ of H~I gas is likely associated with
Phoenix.

Studying star formation in dwarf galaxies has the potential to provide
clues about the mechanisms which govern star formation, and the
relative formation epochs for different types of galaxies. Key
questions include the age of the oldest stellar populations -- are
dwarf galaxies the oldest stellar systems and the building blocks for
larger galaxies? -- and the degree to which star formation is
episodic.  The latter issue has implications for the visibilities of
different types of galaxies at high redshifts; if star formation is
strongly episodic, then even small galaxies can appear bright when
observed at redshifts for which star formation rates were greatest.

We initiated a HST/WFPC2 study of Phoenix to investigate its stellar
population in greater detail. In particular, our project was motivated by
the desire to determine whether an ancient population of stars exists by
looking for a horizontal branch population.  During the period between
proposal submission and the time the observations were taken, however,
several groups detected such a horizontal branch from ground-based
observations (Held \etal 1999, Martinez-Delgado \etal 1999, Tolstoy \etal
2000). Nonetheless, the deeper HST data still provide several pieces
of valuable information about the star formation history of Phoenix,
which we discuss in this paper:  1) WFPC2 photometry of the horizontal
branch is more precise than can be obtained from the ground, leading
to the ability to study its morphology and determine its magnitude with
higher precision, 2) the deeper observations allow us to observe stars
down to the level of a turnoff for an ancient population, providing
information on the distribution of ages of stars in Phoenix, and 3)
accurate photometry of the younger main sequence stars allows us to
assess the degree to which more recent star formation has been episodic
and also to get some indication of the relative metallicities of the
young population compared with the older one.

In section 2 we discuss our observations and data reduction. Section 
3 discusses the morphology of the color-magnitude diagram. 
In Section 4 we use numerical simulations of the distribution of stars in the
color-magnitude diagram to infer the star formation history of Phoenix, and
discuss the uncertainties which attend such an analysis. Some implications
of our Phoenix results for the evolution of dwarf galaxies in general are 
discussed in Section 5.

\section{Observations and Reductions}

One field in Phoenix centered on $\alpha$ (J2000) = 01:51:06.3, $\delta$
(J2000) = -44:26:40.9 was imaged with WFPC2 on HST on 15 January 1999.
While appearing to be close to the center of the galaxy as seen on the
digitized POSS, this field was chosen to avoid bright (field) stars. Our
program was specifically intended to concentrate on the old background
stellar population of Phoenix.  However, given that various authors
(Held \etal 1999, Martinez-Delgado \etal 1999, Ortolani \& Gratton
1988, Canterna \& Flowers 1977) find the youngest stars in Phoenix to
be concentrated towards the center of the galaxy, our WFPC2 field does
sample these stars as well. Table 1 shows the relevant
parameters for the HST observations.

The WFPC2 observations were obtained over a total of 4 orbits. Six
exposures were made through each of the F555W and F814W filters. Half
of the exposures were made at one telescope pointing, while the
remaining half were slightly offset by $\sim 0.25$ arcsec along the
axes of the WFPC2 CCDs. Data were processed through the STScI pipeline.
The combined exposures in F555W at one of the pointings are shown in
Figure \ref{fig:phoenix}.

Photometry was obtained using simultaneous PSF-fitting on the entire
stack of frames, as described in Holtzman \etal (1997), using custom
software within the XVISTA data reduction package that was developed/modified 
based on the routines of Stetson (1987); this software
has been extensively used to reduce HST/WFPC2 data. In brief, all
frames were constrained to have the same lists of stars with the same
relative positions (allowing for slight scale changes between the two
different filters). The stars in different exposures through the same
filter were constrained to have the same relative brightnesses, with
frame-to-frame scaling constrained by the relative exposure times.  Model PSFs
were derived individually for each frame using phase retrieval
of a few bright stars to derive the mean focus for each frame, 
individual exposure jitter information, and the most recent 
estimates for the variations of aberrations and pupil functions across
the WFPC2 field of view.  Prior to the PSF fitting, frames at each of
the two pointings were compared with each other to flag cosmic rays in
each frame, and these were ignored during the fitting process.
Aperture corrections were determined by comparing the PSF magnitudes
with measurements of stars through $0.5\arcsec$ radius apertures made
on frames in which all neighbors had been subtracted using the
PSF-fitting results.  Both PSF-fitting and aperture photometry results
were corrected for the effect of charge transfer efficiency problems
in the WFPC2 detectors using the prescription of Dolphin (2000); results
presented in this paper were largely unaffected by this correction, however.

Zeropoints from Holtzman \etal (1995) were used
to place the instrumental magnitudes on the synthetic WFPC2 photometric
system (which we use for comparison with isochrones), and
transformations to Johnson/Cousins V and I from Holtzman \etal (1995)
were used for comparison with previous results on the tip of the red
giant branch and giant branch colors.

The V vs. (V-I) and I vs. (V-I) color-magnitude diagrams (CMDs) for the
full WFPC2 field observed in Phoenix are shown in Figure
\ref{fig:cmd}.  Photometric errors as estimated from the PSF-fitting
are plotted versus $V$ magnitude in Figure \ref{fig:err} and
demonstrate the high accuracy of the WFPC2 photometry; typical
1$\sigma$ photometric errors at the horizontal branch are $\ltaprx
0.02$ mag.

More detailed photometric errors were derived using a set of twenty
artificial star tests. In each test, over a thousand simulated stars of
a given brightness were added to images in a grid pattern (to prevent
additional crowding) and photometry was rederived for all stars on the
frame. The 20 tests covered brightnesses between $20<V<28$, with finer
sampling at the faint end to measure the completeness of the
photometry; at each brightness, the artificial stars were chosen to
have the median color of the observed stars at that brightness.
Photometry for the simulated stars at each magnitude were
extracted to provide error histograms. These were used in the
construction of artificial CMDs discussed below. In general, the true
errors (difference between observed and simulated brightnesses) are
comparable to those estimated from the PSF-fitting routines for the
bulk of the simulated stars, but there is always a subset of stars
which have significantly larger errors. These are generally the product
of systematic errors caused by crowding, and as a result, are
correlated in the two bandpasses. In addition, the errors for the
artificial stars are smaller than those estimated by the PSF-fitting
routines for the brighter stars, which is most likely a result of using
the exact same PSF to create and reduce the artificial stars.  An
indication of these effects is shown in the bottom panel of Figure
\ref{fig:err}; the various vertical bands are the measured errors for
the simulated stars.  The distribution of these errors is non-Gaussian
since a small subset of stars have relatively large errors; still, it
is apparent that the bulk of the artificial stars have observed errors
consistent with the rms errors which are output from the PSF-fitting
routine (and shown as the continuous set of points). We emphasize,
however, that we use the observed error distribution from the artifical
star tests in our detailed comparison of simulated CMDs with the data
discussed below.

Error bars plotted in Figure \ref{fig:cmd} show the rms errors from
the artificial star tests as a function of magnitude. Since the true
distribution of errors is non-Gaussian, these do not fully convey all
the information we have used about the error distribution; the error
bars shown in Figure \ref{fig:cmd} were computed by determining an rms from
the measured error distribution with rejection of $\pm 3\sigma$ outliers.

\section{The Color-Magnitude Diagram} 

The CMD of Phoenix provides examples of stars in
many different stages of stellar evolution.  The CMD exhibits the types
of sequences, most notably the near-vertical blue and red ``plumes,''
which are typical of dwarf irregular galaxies (\eg, Cole \etal 1999).
The main stellar sequences which constitute such plumes are illustrated
by Dohm-Palmer \etal (1997) and Gallart \etal (1996).

The vertical blue sequence is the main sequence of a young stellar
population in Phoenix. The red sequence is a combination of giant branch
and asymptotic giant branch stars which can have a wide range of ages and
metallicities.  In addition, there is a clear detection of a horizontal
branch in Phoenix, with both blue and red components.  There is a red clump
of stars found near the intersection of the horizontal branch and the
giant branch that extends more than 0.5 mag in V and I brighter than the
red horizontal branch (RHB). The stars in this clump that are brighter
than the RHB are intermediate-age (1-10 Gyr) core helium-burning stars.
Finally, there are indications of two sequences extending upwards from the
red clump; the most populated one is nearly vertical, and is particularly
pronounced in the magnitude range $22 < V < 24$, while a second very
sparsely populated sequence gets bluer at brighter magnitudes; these
(especially the bluer one) can perhaps be seen more clearly in the larger
area ground-based observations of Held \etal (1999). These two sequences
correspond to core helium burning stars at either end of the blue loops
through which relatively young massive stars evolve.

\subsection{Red giant branch}

The red giant branch (RGB) is located between $V-I = 0.7$ and 1.6, 
and extends brightward to $I\sim 19.1$. The color of the RGB can
be used to place constraints on the metallicity of stars in Phoenix,
and the brightness of the tip can be used as a distance indicator.

If we adopt a distance modulus of $(m-M)_0=23.1$ (Held \etal 1999, and see
below), then the mean color of the red giant branch at $M_I = -3.0$
($I=20.1$) is $V-I \approx 1.22$, or $(V-I)_0 \approx 1.19$.  Using the
calibration of Da Costa \& Armandroff (1990) between $(V-I)_0$ at $M_I =
-3.0$ and [Fe/H], gives a mean metallicity for the Phoenix red giants of
[Fe/H] $\sim -1.87$.
Our estimate of the
mean metallicity of Phoenix is in good agreement with the result of Held
\etal (1999), $\sim 0.5$ dex lower than that of Martinez-Delgado \etal
(1999), and 0.13 dex higher than that of van de Rydt \etal (1991).
With an absolute visual magnitude of $M_V = -9.8$ (Pritchett \& van den
Bergh 1999), Phoenix falls very close to the [Fe/H] versus
$M_V$ relation obtained by Caldwell \etal (1992) for Local Group dwarf
elliptical and spheroidal galaxies.

The giant branch at $M_I = -3.0$ is $\sim 0.15$ mag wide in $V-I$. If
this is assumed to be entirely due to a metallicity spread within an
ancient stellar population then the calibration of Da Costa \&
Armandroff (1990) indicates a metallicity spread within Phoenix of
[Fe/H] $\sim -1.45$ to $-1.9.$ However, two effects complicate the
above determination of the mean metallicity and metallicity spread of
Phoenix from giant branch colors. First, the blue edge of the giant
branch is likely to include some asymptotic giant branch stars. Second,
and probably more important, the giant branch color is a function of
age as well as metallicity, in that younger giants are bluer than older
ones. At fixed metallicity but mixed ages, the giant branch color
derived from assuming an exclusively old population will underestimate
the mean metallicity, and the giant branch width will overestimate the
metallicity spread. However, if ages and metallicity are correlated, as
would occur if younger populations are enhanced in metals, the true
metallicity spread can be significantly \textit{larger} than that
inferred from the giant branch width assuming an exclusively old
population.

Although there is some uncertainty about the metallicity depending on the
age, there is still an upper limit which can be placed on the metallicity
of any component of the population.  Even if the entire population was
young (less than a few Gyr), the color of the giant branch would still
require a metallicity less than [Fe/H] $\sim$ -0.7 based on recent Padova
isochrones (Girardi \etal 2000, Girardi \etal 1996).  This is
demonstrated in Figure \ref{fig:isorgb}, which shows the Phoenix CMD
with isochrones for ages 1.6, 3.2, 6.4, and 12.8 Gyr at $Z=0.0004$ and
$Z=0.004$.  In order for the RGB to be consistent with a metallicity of
Z=0.004 ([Fe/H] = -0.7), the Padova isochrones indicate that the
ages of RGB stars would have to be less than 2 Gyr.

The deep HST data allow us to constrain the population further, since
the red giant branch of Phoenix is found to be well populated right
down to $I \sim 25.8$, \ie, down to the luminosities expected for the
base of the RGB of a 13 Gyr $Z=0.0004$ stellar population. This
demonstrates conclusively that not all the stars are young.  Within the
allowed range of metallicity, giants which are this faint must be older
than $\sim$ 6 Gyr, based on comparisons with the Padova isochrones; for
such ages, the dependence of giant branch color on age is relatively
weak. Since a significant fraction of the red giants must be relatively
old, the isochrones indicate that they must also be relatively
metal-poor, \ie [Fe/H] $\ltaprx$ -1.7.

A more quantitative discussion of the metallicity spread within Phoenix
requires analysis of the relative numbers of younger and older stars.
This is discussed further below.  However, in terms of both the mean
metallicity and the extension to faint magnitudes, the RGB of Phoenix
is consistent with this galaxy containing a substantial component of
stars analogous to Population II of the Milky Way.


The magnitude of the tip of the red giant branch can be used as a
distance indicator for metal-poor populations.  The accuracy of this
technique in the current data set may be limited by the relatively
small number of stars; one might expect that the tip magnitude derived
from the relatively smaller number of stars in the WFPC2 frame would be
fainter than a true tip magnitude, since our tip is essentially defined
by the single brightest red giant. Still, we find an apparent tip
magnitude of $I_{TRGB} = 19.1$, which is essentially identical to the
tip magnitude derived by Held \etal (1999) and just $\sim$ 0.1 mag
fainter than the RGB tip determined by Martinez-Delgado \etal (1999),
both from ground-based photometry.  We adopt an absolute magnitude of
$M_{I,TRGB} = -4.05$ for the tip (Lee \etal 1993) and a reddening of
$E(B-V) = 0.02$ (Burstein \& Heiles 1982), giving E(V-I) = 0.026
($A_V=0.062, A_I=0.036$). This yields a distance modulus for Phoenix
based on the tip of the RGB of $(m-M)_0 = 23.11$; the error in this
quantity is several percent based on uncertainties in the adopted
extinction. If the true tip is brighter than our brightest star, the
distance modulus would be smaller. Our distance provides additional
support for the derived TRGB distances of Held \etal ($(m-M)_0=23.04$)
and Martinez-Delgado \etal ($(m-M)_0=23.0$).

\subsection{Horizontal branch}

A horizontal branch (HB) can be discerned at $V = 23.9$ and $0.0 < V-I
< 0.8$ which merges with the red giant branch at $V-I \sim 0.8$.  There
are well-defined blue (BHB) and red (RHB) components of the horizontal branch,
separated by a gap which likely contains RR Lyrae stars.  At first
glance the occurrence of a RHB in Phoenix, which has a mean metallicity
as low as [Fe/H] $\sim -1.8$ (\ie, comparable to Galactic globular clusters
such as M13 and M3), might indicate that Phoenix is an example of a
second-parameter stellar system. However, if there is a metallicity spread, at
least some of these RHB stars may have evolved from the most metal-rich
old red giants in Phoenix. Given that there is likely both an age and
metallicity (see below) spread in Phoenix, interpreting the morphology
of the horizontal branch is a complex problem.  Nevertheless, the
presence of a distinct blue HB clearly indicates that there is an
ancient population of stars in Phoenix, as noted by Tolstoy \etal
(2000).

The accurate WFPC2 photometry allows a measurement of the horizontal
branch magnitude at $V=23.9\pm0.1$. From ground-based photometry,
Held \etal (1999) estimated the horizontal branch to be at $V=23.78\pm
0.05$. Using the Lee \etal (1990) relation between the absolute visual HB
magnitude and metallicity, our measurement gives a distance of $(m-M)_0
= 23.3$, larger than the TRGB method, but perhaps not in bad
agreement given uncertainties in the HB calibration.

\subsection{Main sequence}

A young main sequence locus extends brightward to $V \sim 21.5$ (with a
few even brighter stars) or $M_V  \sim -1.6$.  A similar blue plume
with $-0.5 < V-I < 0.2$ was found by Martinez-Delgado \etal (1999) at
magnitudes brighter than $I \sim 23.5$ (which corresponds to the faint
limit of their photometry at $V-I \sim 0.0$), and is evident in the
region $V < 24.0$ and $-0.2 < B-I < 0.0$ of Fig.~12 of Held \etal
(1999).  In these ground-based CMDs the main sequence population does
not appear to be as prominent as seen in the WFPC2 CMD; this may be an
effect of larger photometric errors smearing out the ground-based
observations of the main sequence and increasing incompleteness for
bluer stars in the ground-based observations. The accuracy of the WFPC2
photometry shows that the main sequence is well populated, and
hence that the young star contribution to the total population of
Phoenix is significant, as will be shown quantitatively below.

Figure \ref{fig:iso} shows the Padova isochrones for two ages (100 Myr
and 3.2 Gyr) and metallicities of $Z=0.0004$ and 0.004. The
extension of the main sequence up to $M_V\sim -2$ suggests that stars
as young as 100 Myr exist in Phoenix, in good agreement with previous
works (Martinez-Delgado \etal 1999, Held \etal 1999).  Main sequence
turnoff stars this bright would have masses of $\sim 5.0$ $M_\odot$.

Several points are apparent from Figure \ref{fig:iso}. First, there are
many stars which fall significantly redward of any of the ZAMS
isochrones. The observed width (in color) of the upper main sequence is
significantly larger than expected from observational errors; from
artificial star tests we find that magnitudes for the bulk of the stars
are measured to an accuracy of a few percent down to $M_V\sim 2$.  The
color spread of main sequence stars almost certainly implies that stars
with a range of ages exist in Phoenix; it is highly unlikely that the
color spread results entirely from metallicity spread, since very
metal-rich stars would be required. Such a metal-rich population would
produce giants significantly redder than any that are observed.  Stars
which fall redward of the ZAMS are stars which are evolving off of the
main sequence.  Despite the accurate photometry, there are no
pronounced main sequence turnoffs in the CMD between ages of 100 Myr
and 4 Gyr, suggesting that the young population in Phoenix is
\textit{not} the result of one or several distinct bursts of star
formation, but rather is the result of a more continuous star formation
history over the past several Gyr. This is consistent with the analysis
of Martinez-Delgado \etal (1999), who estimate that the recent ($<1$ Gyr) star
formation rate is comparable to the average star formation rate for the
age interval 1-15 Gyr.

It appears that the $Z=0.0004$ isochrone ([Fe/H] = $-1.7$, \ie the
metallicity inferred for the giants) falls blueward of the main body of
the main sequence, suggesting that the younger stars in Phoenix have
significantly higher metallicities than the older population. This
would imply that Phoenix has undergone chemical enrichment over its
lifetime.  This suggestion is supported by the giant branch colors as
well; if all of the stars in Phoenix had [Fe/H] $\sim$ -1.7, the younger
giants would be bluer than any of the observed giants.

\subsection{Other core helium burning stars}

A clear red clump of stars exists near the intersection of the
horizontal branch with the red giant branch. These are intermediate age
(several Gyr) core helium burning stars, and confirm the existence of
an intermediate age population suggested by the width of the main
sequence.

A sequence of red helium-burning stars (with ages of around
0.4-0.8 Gyr) extends vertically upwards from the red end of the
horizontal branch to an $I$ magnitude comparable to the tip of the red
giant branch. This sequence is located about 0.2 mag blueward of
the $Z=0.0004$ RGB isochrones. These are younger stars in the reddest
stage of the so-called ``blue loops'' of later stellar evolution. 
There is a hint of a sequence corresponding to the blue end of the blue loops; 
this sequence slopes towards the main sequence and 
although it has only a few stars, 
it is defined by the observation that there are gaps around it in the
CMD, suggesting that these stars are not foreground stars.
The exact details of the ages of these stars is difficult to ascertain
without knowledge of their metallicities.

Intermediate-age (1-10 Gyr) asymptotic giant branch stars are seen in
CMDs of dwarf irregular galaxies such as IC 1613 (Cole \etal 1999),
Sextans A (Dohm-Palmer \etal 1997), and NGC 6822 (Gallart \etal 1996), 
where they extend redward of $V-I
\sim 1.6$ from the region of the tip of the first ascent giant branch.
There are few such stars apparent in the Phoenix CMD of Figure
\ref{fig:cmd}.  However, Held \etal (1999) found a significant number
of such stars in their ground-based CMD and they concluded on this
basis that intermediate-age stars constitute a substantial component of
Phoenix; additional evidence for the presence of AGB stars is discussed
by Martinez-Delgado \etal (1999).

\section{The Star Formation History of Phoenix}

Given that stars of many different ages are present in Phoenix, it is
of interest to know the relative number of stars at different ages, or
alternatively, the history of the star formation rate. To some extent,
the answer to this question depends on where one looks in Phoenix,
since it is apparent from larger field studies (Held \etal 1999,
Martinez-Delgado \etal 1999) that the star formation history varies
across the galaxy, with younger stars being more centrally
concentrated. Here we consider somewhat more quantitatively the
approximate star formation history that is required to match the
populations seen in the central region of Phoenix imaged by HST/WFPC2.
We can also try to quantitatively address the relative numbers of
stars of different metallicities.

There are many complications to extracting star formation histories
from the distribution of stars in a CMD, and several groups have
developed different methods for doing so (\eg, Tolstoy \& Saha 1996;
Dolphin 1997; Hernandez, Valls-Gabaud, \& Gilmore 1999; Ng 1998;
Gallart \etal 1999; Holtzman \etal 1999). In the current paper, we
analyze our CMD based on the methods of Holtzman \etal (1999), in which
the number of stars in many regions of the CMD are fit using
least-squares techniques to a combination of basis functions of stellar
populations of different ages and metallicities. The best fit was
derived using the $\chi_\gamma^2$ estimator of Mighell (1999).

The interpretation of derived star formation histories, and in particular
an understanding of the uncertainties in the results, is a complex issue.
Detailed comparisons of results obtained by using different methods have
not yet been performed, and the sensitivity of the results to
systematic errors, \eg, in the input stellar models or assumed initial
mass function (IMF), is difficult to assess.  Consequently, the results
presented here should be viewed as preliminary; we attempt to assess
uncertainties by showing results not only for different parameters, but
also using different subsets of the observed data to derive star
formation histories.

For the current purpose, we derive a star formation history assuming an
IMF with the Salpeter (1955) slope. Since we do not sample the lower
main sequence of completely unevolved stars, independent constraints on
the IMF slope are difficult. We also assume that unresolved binaries
are not important in the observed CMD of Phoenix, which will be the
case regardless of the true binary fraction in the galaxy if the masses
of binary stars are drawn independently from the same initial mass
function. This is because the most likely companion masses in this case
are significantly lower than the masses of the stars which are observed
and thus do not strongly affect the system luminosity or color. The
issue of whether masses in binary star pairs are uncorrelated or not is
still the matter of significant debate.

We derive star formation histories for different age bins, where the
width of the age bins increases logarithmically with lookback time.
This is necessitated by the fact that it becomes increasingly difficult
to resolve (in a temporal sense) star formation events as one goes to
older ages, since isochrones become more and more similar for older
ages. As a result, the derived star formation histories cannot provide
information about the variation of the star formation rate on time
scales shorter than the width of the age bins, roughly 25\% of the age
of any given population. Within each age bin, the star formation rate
is assumed to be constant.

Star formation histories were extracted from the Phoenix CMD for five
different choices of distance modulus: $(m-M)_0 = 22.9$, 23.0, 23.1,
23.2, and 23.3.  Using $(m-M)_0 = 23.1$, the distance as derived from
our TRGB, gave the best fits (lowest $\chi^2_\gamma$), although they
were only marginally better than those using $(m-M)_0=23.0$.  For the
reasons discussed above, it was not possible to find good fits to the
CMD using an exclusively metal-poor population, or for that matter,
using any star formation history in which the metallicity does not
change over time.  This is because the giant branch requires a
metal-poor population, while the main sequence requires a population of
higher metallicity. As a result, we imposed no constraints on an
age-metallicity relation to fit the Phoenix population; the constituent
basis stellar populations cover all combinations of age and
metallicity, although only seven discrete metallicities (corresponding
to those for which we have isochrones from the Padova group) were used.

Results for our best-fitting model are shown in Figure \ref{fig:sfh}.
The upper left panel gives the best fitting star formation rate as a
function of lookback time. Since we only cover a small region of the
galaxy, this star formation history is given in a relative sense only;
the total number of stars formed is normalized to sum to unity.  The
quality of the fit can be judged by looking at the residual Hess
diagram shown in the lower right. This panel shows the difference
between the number of observed and the number of stars in the
best-fitting model, in units of the expected $1\sigma$ errors from
counting statistics; the diagram is scaled so that the range of black
to white covers from $-3\sigma$ to $+3\sigma$, where $\sigma$ is
defined based on the suggestions made by Mighell (1999).  Other panels
show the derived cumulative age and metallicity distribution functions,
and a comparison of the observed and model luminosity functions. In
general, results are more uncertain for the youngest populations ($< 1$
Gyr), since the stars that are uniquely contributed by these
populations  (upper main sequence stars) comprise a relatively small
number of the stars in the color-magnitude diagram.

The lower left panel gives a representation of the ``population box''
(Hodge 1989) for the central regions of Phoenix. This combines
information about the derived formation rate as a function of age and
metallicity.  Remarkably, without imposing any constraints on the
age-metallicity relation, we recover a rough relation in which
metallicity increases with time; a similar result also emerged from a
similar analysis of several fields in the LMC (Holtzman \etal 1999).
This lends quantitative support to our suggestion that the metallicity
in Phoenix has evolved over time; the derived metallicity distribution
is shown in the upper right panel.  The population box may also
highlight some problems with the models as well; the observed
color-magnitude diagram appears to require a small population of young
metal-poor stars. The presence of this feature comes from the existence
of a few relatively blue upper main sequence stars, so it is difficult
to know for sure how seriously to consider them; accurate photometry
over a larger field is required.

A comprehensive understanding of the uncertainties in the results
presented in Figure \ref{fig:sfh} is difficult. The formal errors on
the derived star formation rates are relatively small, usually 5-10\%
for most of the age bins (larger for the youngest ages).  However, the
true uncertainties are probably significantly larger and systematic,
depending on the choice of constituent basis stellar populations,
distance, reddening, IMF, stellar models, the use of discrete
metallicities, and the details of how the best fit is defined.  Our
experience with allowing these to vary suggests that while the star
formation rate in individual age bins can vary significantly (factor of
two), the variations are correlated such that changes in derived star
formation rate for one stellar population are generally accompanied by
nearly opposite changes in other similiar (\ie, in age or metallicity)
populations. As a result, the cumulative age distribution tends to be
relatively robust against systematic errors. This is demonstrated in
Figure \ref{fig:sfhcum}, which shows cumulative age distributions for a
variety of models. The solid curves all use the same distance modulus,
but are determined using different subsets of the observed data:
results were derived using two different faint magnitude cutoffs
($M_V\sim 2.9$ and $M_V\sim 3.5$) for the region of the CMD which was
fit and using both the complete data set as well as using randomly
selected subsamples of half the observed stars. The dotted curves show
the systematic changes resulting from using different distance moduli,
although the extremes of these choices ($(m-M)_0=22.9$ and 23.3)
provide noticeably poorer fits.

Our overall conclusion is that the CMD of Phoenix requires star
formation to have been ongoing at a roughly constant rate over a Hubble
time. The fits suggest that the bulk of the stars even in the central
region of Phoenix are older than $\sim$ 5 Gyr.

The details of fluctuations in the overall star formation rate are
significantly more uncertain.  Although the fit in Figure \ref{fig:sfh}
suggests some fluctuations in the star formation rate, we feel that the
actual evidence for strong variations is weak, although they are
certainly not ruled out.  Different choices of population parameters
(\eg, distance) or what regions of the CMD are fit can change the timing
and amplitude of these fluctuations. As a result, it is less clear as
to what the current results show about whether the star formation rate
in Phoenix is episodic. However, it is clear from an inspection of the
CMD that there are no distinct main sequence turnoffs as can
be seen, for example, in the CMDs of the Carina dwarf spheroidal
(Smecker-Hane \etal 1996).

One of the main reasons for uncertainties in the details of the star
formation history comes from the age-metallicity degeneracy of the
location of stars around the turnoff in a CMD; different combinations
of age and metallicity can place stars at the same location in the
CMD.  Without the presence of the unevolved lower main sequence in the
current data or independent measurement of metallicities, it becomes
difficult to accurately separate these different populations.  The
situation would be improved with significantly deeper photometry and/or
with the measurement of independent metallicities for individual stars
in Phoenix.

\section{Discussion}

Deep HST/WFPC2 imaging has provided a color-magnitude diagram (CMD) of
the central regions of the Phoenix dwarf galaxy at unprecedented
accuracy.  The presence of stars at all ages is apparent in this
diagram. A distinct horizontal branch and the presence of a red giant
branch which extends all the way down to a turnoff expected for an old
population indicates that this system has a substantial component of
metal-poor stars with metallicities comparable to the mean of the
Galactic halo; Phoenix appears to have an underlying component of
ancient Population~II stars. In this respect it is similar to dwarf
spheroidal galaxies like Ursa Minor, Draco, and Sculptor, as previously
concluded by Held \etal (1999), Martinez-Delgado \etal (1999), and
Tolstoy \etal (2000).

However, the existence of red clump stars and main sequence stars
confirm that Phoenix has experienced considerable star formation over
the past few Gyr, with star formation possibly as recent as 100 Myr
ago. The location of the main sequence suggests that the metallicity in
Phoenix has evolved with time; old blue horizontal branch stars and the
giant branch stars require low metallicities, comparable to those in
the halo of the Milky Way, while the younger main sequence stars
suggest somewhat higher metallicities.

Fits to the observed CMD suggest that star formation has been roughly
continuous over the lifetime of Phoenix.  There is no obvious evidence
for strongly episodic star formation, although the formal fits for the
star formation history suggest that a mildly varying star formation
rate can certainly fit the data.  One caveat to this statement is that
it applies only over relatively long (\eg, Gyr) timescales; burstiness
on very short timescales would be essentially impossible to distinguish
from the CMD for all but the most recent past. Another caveat is that
the details of the derived star formation rate are suspect due to a
variety of reasons including the age-metallicity degeneracy;
independent measurements of metallicities of stars in Phoenix should
improve the situation considerably.

The implications are that a dwarf galaxy can have star formation which
extends over nearly a Hubble time.  With roughly continuous star
formation, Phoenix appears to resemble the Magellanic Clouds (\eg,
Holtzman \etal 1999). However, Phoenix is a significantly less massive
system in which it might be expected that the loss of interstellar gas
and interruptions in star formation would be much more important. In
fact, the apparent star formation history of Phoenix stands in contrast
to the strongly episodic star formation of Carina (Smecker-Hane \etal
1994, 1996, Hurley-Keller \etal 1998), a dwarf galaxy that is only 0.8
mag fainter than Phoenix (Mateo 1998).  It appears that the mechanisms
governing star formation in dwarf galaxies may be rather complex.

The HST/WFPC2 data on Phoenix add support to the growing
evidence in the literature that dwarf galaxies, both dIrrs and dSphs,
can have very varied star formation histories, particularly in recent
epochs. By contrast, it may be possible that there is at least some
commonality between dwarf galaxies in regard to their early chemical
evolution. As noted in Section 3.1, Phoenix appears to follow the same
integrated-magnitude versus metallicity relation as dwarf spheroidal
galaxies in the Local Group, some of which, like Draco and Ursa Minor,
do not appear to have supported recent star formation such as that of
Phoenix. Nonetheless the degree of chemical enrichment in these
systems, as indicated by the color of the old giant branch, would
appear to be correlated with the overall (luminous) mass of the
galaxy.  This suggests that the amount of metal enrichment during the
earliest phases of star formation within dwarf galaxies (perhaps more
than $\sim 10$ Gyr ago) does vary in a way that depends to some extent
on basic galaxy properties such as mass. In classical closed box
models of chemical evolution the average metallicity of all stars
formed up to a given time is proportional to the fraction by mass of
the galaxy that has turned into stars (\eg, Smith 1985). Within the
context of such a model, the observations indicate that the fraction of
a galaxy which is turned into old metal-poor stars, and the attendant
early chemical evolution history, is governed by similar mechanisms in
both dSph and dIrr galaxies.

Support for this work was provided by NASA through grant number GO-0698.02-95A
from the Space Telescope Science Institute, which is operated by the Association
of Universities for Research in Astronomy, Incorporated, under NASA
contract NAS5-26555.

\newpage

\begin{table}[h]
\begin{center}
\begin{tabular}{c c c}
\multicolumn{3}{c}{\bf Table 1.} \\
\multicolumn{3}{c}{ Summary of Observations} \\ \tableline \tableline
Dataset&Filter&Exposure time (s)\\
\tableline
u48i0201r&F814W&100\\
u48i0202m&F814W&1300\\
u48i0203r&F814W&1000\\
u48i0204r&F555W&100\\
u48i0205r&F555W&1300\\
u48i0206r&F555W&1100\\
u48i0207r&F555W&100\\
u48i0208r&F555W&1300\\
u48i0209r&F555W&1000\\
u48i020ar&F814W&100\\
u48i020br&F814W&1300\\
u48i020cr&F814W&1100\\
\tableline
\end{tabular}
\end{center}
\end{table}

\newpage

\begin{figure}
\caption{The WFPC2 F555W image of the central regions of the Phoenix dwarf from
combined exposures at one of the two pointings. The observed field is
roughly 2.5 arcminutes square (apart from the Planetary Camera quadrant).
The image is shown as the galaxy would appear on the sky, with North
approximately 119 degrees clockwise from vertical.}

\label{fig:phoenix}
\end{figure}

\begin{figure}
\caption{Observed color-magnitude diagrams in the V and I bandpasses.}
\plotone{cmd.epsi}
\label{fig:cmd}
\end{figure}

\begin{figure}
\caption{1$\sigma$ photometric errors as estimated by PSF-fitting routine. 
In the bottom panel, the vertical bars of points are the errors of
individual stars from the artificial star tests.  These show that
although the PSF-fitting errors are reasonable for the majority of
the stars, a significant subset of stars have larger errors because of
crowding.  The large squares are the 3$\sigma$ rejected means of the
artificial star errors. }
\plotone{err.epsi}
\label{fig:err}
\end{figure}

\begin{figure}
\caption{Observed CMD with isochrones from Girardi \etal (2000);
ages of 1.6, 3.2, 6.4, and 12.8 Gyr for metallicities of Z=0.004 (dotted) and
Z=0.0004 (solid) are shown. Absolute magnitudes and colors
were derived using $(m-M)_0=23.1$, $A_{F555W}=0.062$, and
$A_{F814W}=0.036$.  These CMDs are plotted in the WFPC2 synthetic
system for the most accurate comparison with isochrones, which were
also computed for this system.}
\plotone{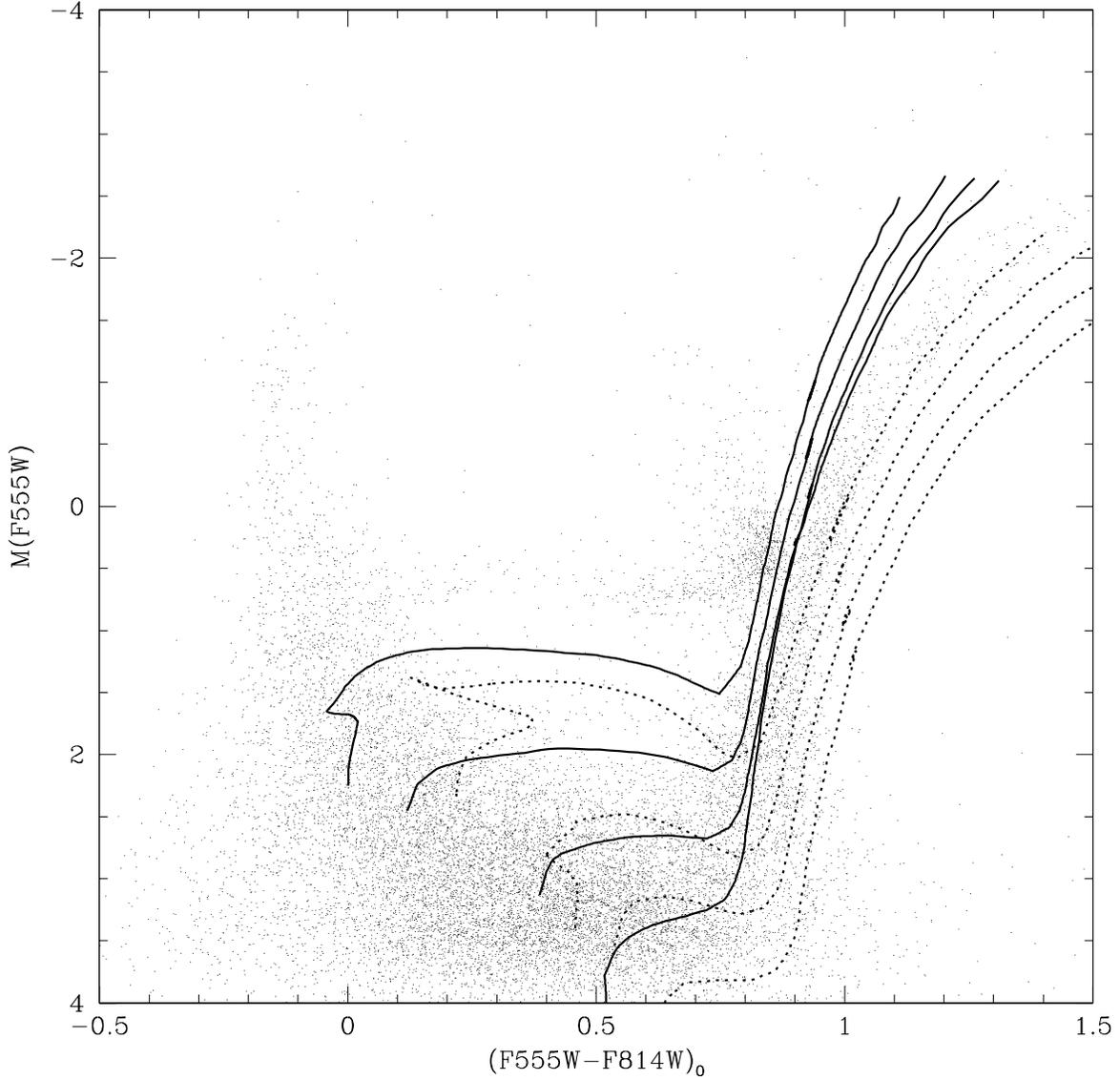}
\label{fig:isorgb}
\end{figure}

\begin{figure}
\caption{Observed CMD with isochrones (100 Myr and 3.2 Gyr) at Z=0.0004 (solid)
and 0.004 (dotted).
Absolute magnitudes and colors were derived using 
$(m-M)_0=23.1$, $A_{F555W}=0.062$, and $A_{F814W}=0.036$. }
\plotone{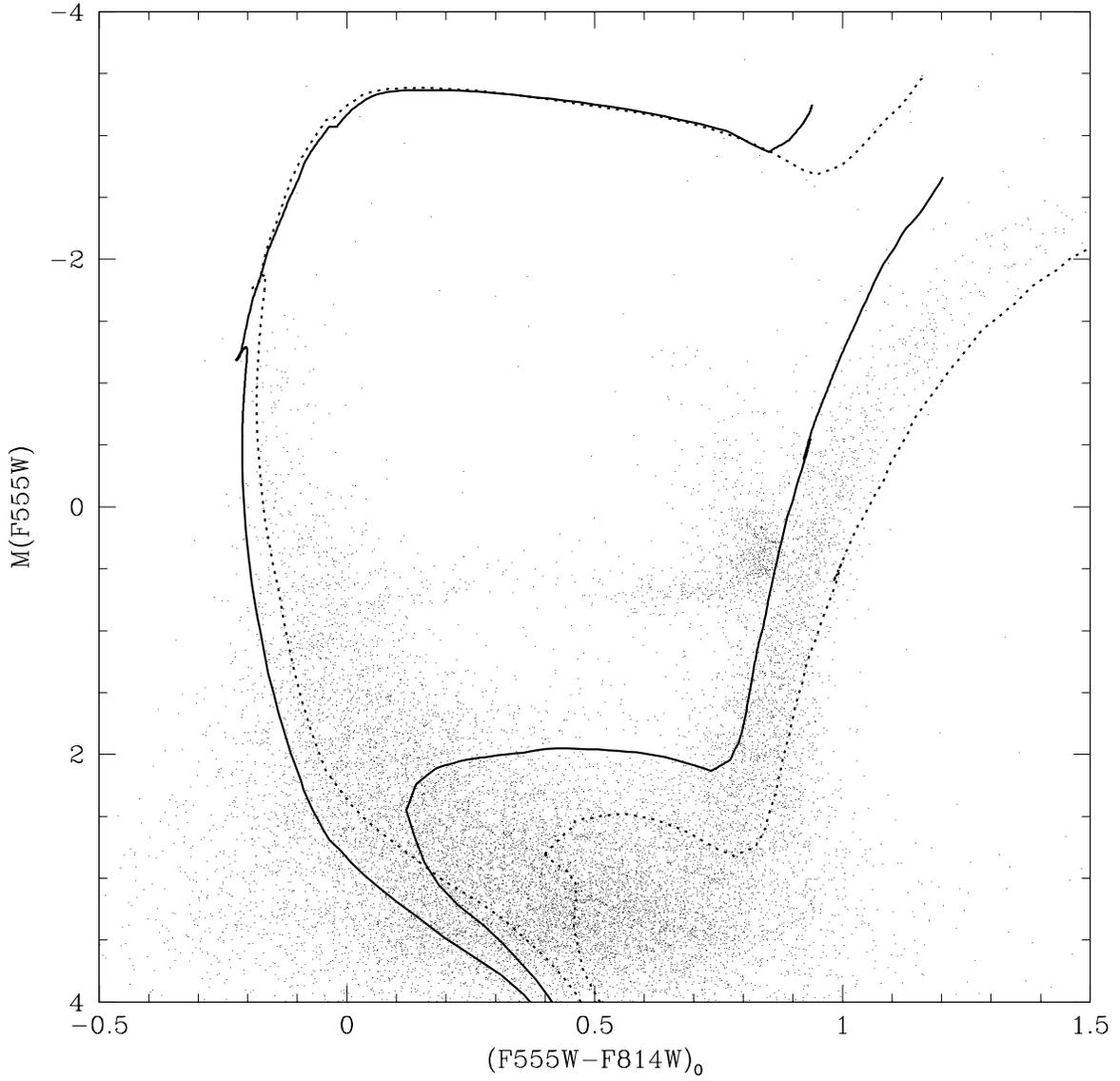}
\label{fig:iso}
\end{figure}

\begin{figure}
\caption{Derived star formation history for central regions of Phoenix,
assuming $(m-M)_0=23.1$. See
text for caveats and discussion of uncertainties. Upper left gives derived
star formation rate as a function of lookback time. Upper right gives
cumulative and differential metallicity distributions; middle left gives
cumulative age distribution. Middle right panel compares observed and model
luminosity functions. Lower left panel shows the derived ``population box'',
the star formation rate as a function of time and metallicity. Lower right
panel shows residuals of data after best-fitting model has been subtracted,
in units of expected variation in each bin in the color-magnitude diagram,
scaled such that black to white ranges from $-3\sigma$ to $+3\sigma$.}
\label{fig:sfh}
\end{figure}
\begin{figure}
\plotone{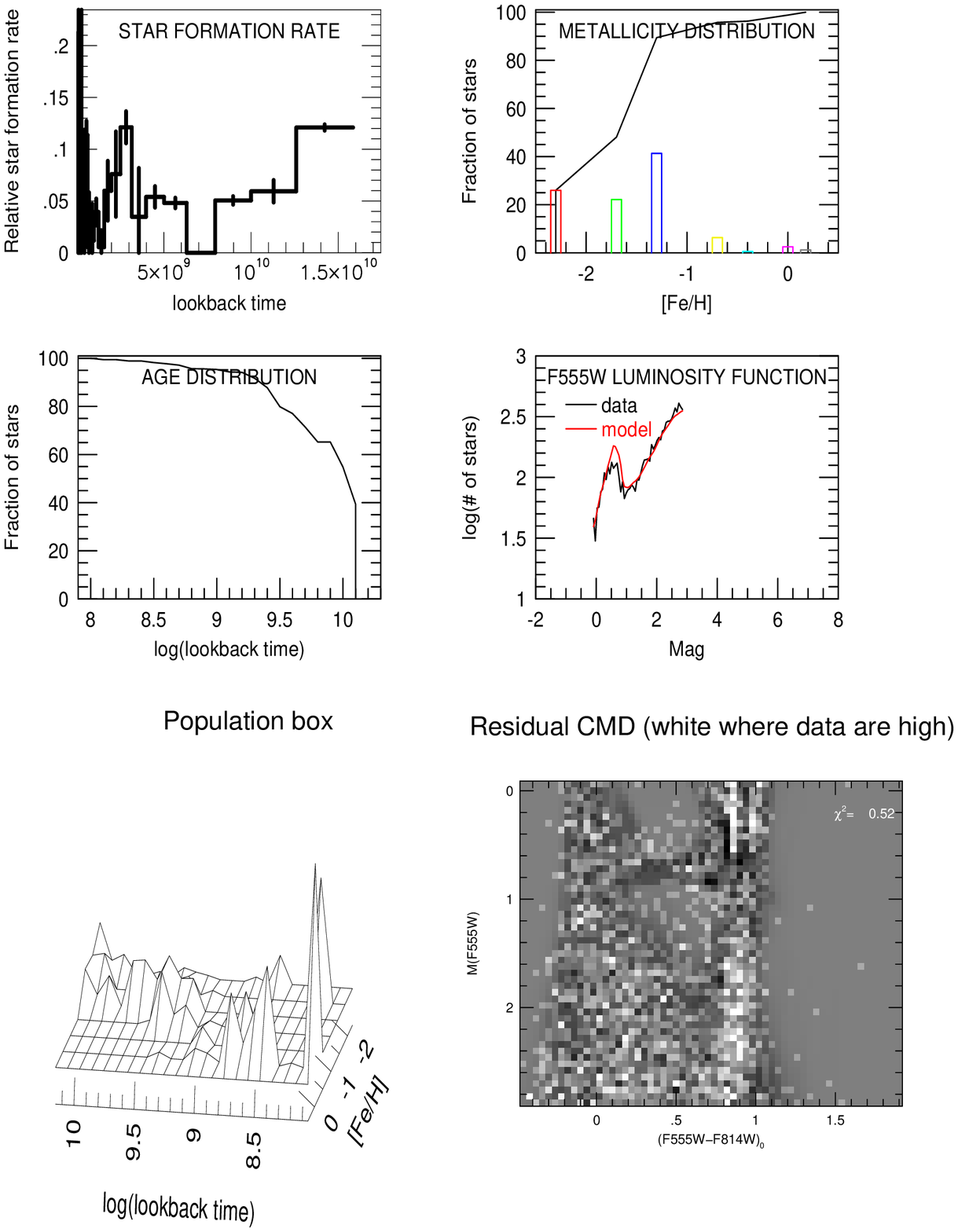}
\end{figure}

\begin{figure}
\caption{Derived cumulative star formation histories for a range of
different fits. Solid curves all assume $(m-M)_0=23.1$, but correspond to
two different faint cutoffs ($M_V\sim 2.9$ and 3.5) in the part of the
color-magnitude diagram which was fit; results are also shown for two
separate halves of the dataset as well as the complete dataset. Dotted
curves show the results for the different distance modulii (22.9, 23.0,
23.2, and 23.3) using the full dataset.} 
\plotone{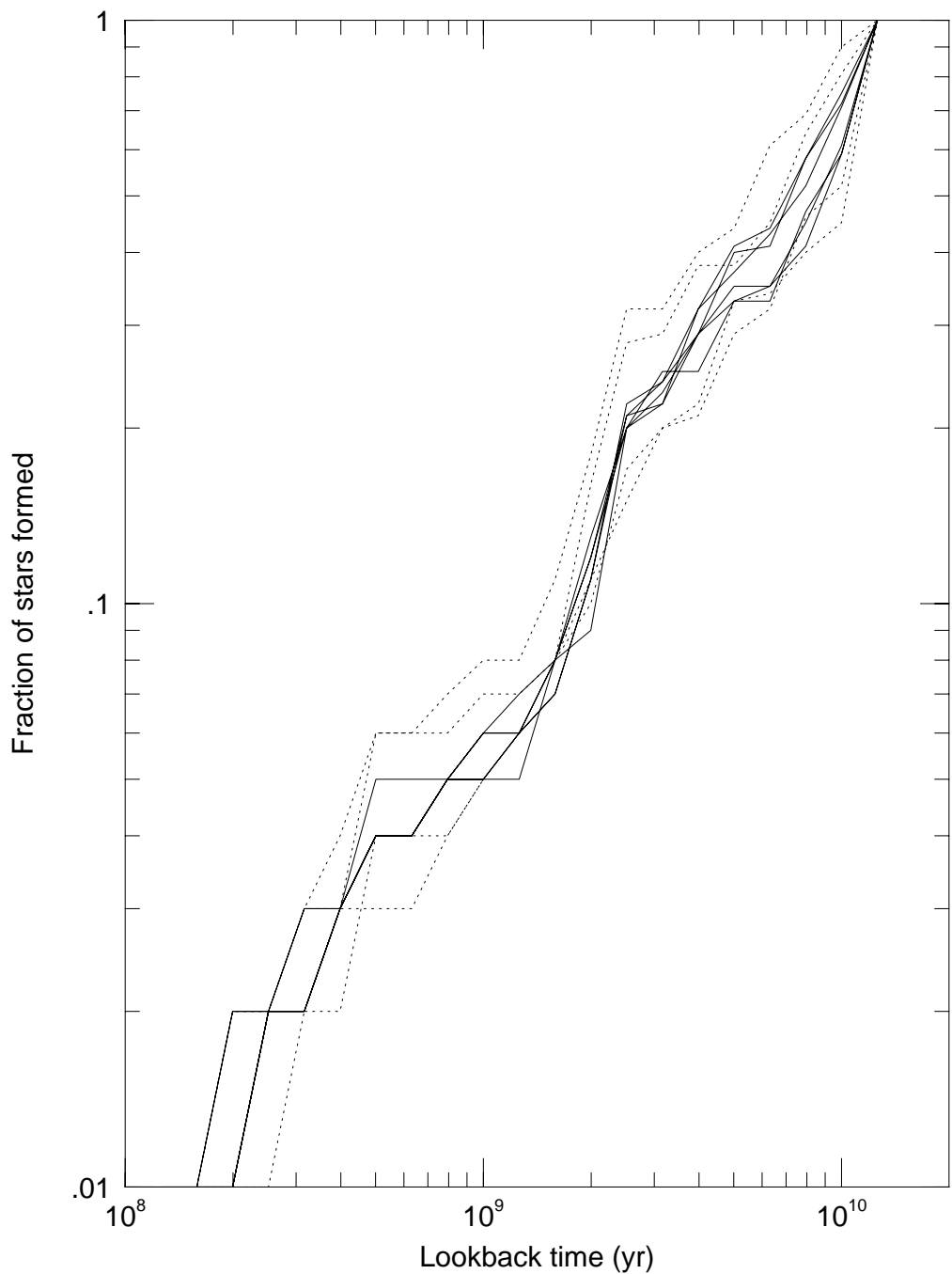}
\label{fig:sfhcum} 
\end{figure}

\end{document}